\documentclass[11pt,a4paper]{article}
\pdfoutput=1
\usepackage{jinstpub}
\usepackage{graphicx}
\graphicspath{{figures_v5/}}
\usepackage{amsmath}
\usepackage{amssymb}
\usepackage{enumitem}
\usepackage{multicol}
\usepackage{multirow}
\usepackage{float}
\usepackage{array}
\usepackage{caption}
\usepackage{subcaption}

\title{Leak Test of Resistive Plate Chamber Gap by Monitoring Absolute Pressure}
\author[a,b]{Suryanarayan Mondal}
\author[b]{V. M. Datar}
\author[b]{Gobinda Majumder}
\author[c]{N. K. Mondal}
\author[b]{K. C. Ravindran}
\author[b]{B. Satyanarayana}
\affiliation[a]{NPD, Homi Bhaba National Institute, Anushaktinagar, Mumbai, India}
\affiliation[b]{DHEP, Tata Institute of Fundamental Research, Homi Bhaba Road, Mumbai, India}
\affiliation[c]{HENPPD, Saha Institute of Nuclear Physics, Bidhannagar, Kolkata, India}

\emailAdd{suryamondal@gmail.com}


\abstract{
  The India-based Neutrino Observatory Project (INO) is a proposed underground high energy physics experiment at Theni, India to study the neutrino oscillation parameters using atmospheric neutrinos. The 50 kton magnetised INO-ICAL detector will require approximately 30,000 of 2m\,$\times$\,2m Resistive Plate Chambers (RPC) as sensitive detectors and proposed to operate for about 20 years. For success of the experiment, each of the RPCs has to function without showing any significant aging during the period of operation. Hence, various tests including a proper leak test are performed during and after production. The methods of leak rate calculation using conventional manometer are valid only when both the volume of the test subject and ambient pressure are kept constant. But both these quantities for a RPC gas gap depend widely on the ambient pressure and temperature. A proper quantitative estimation of the leak rate cannot be acquired from such pressure measurements. By monitoring the absolute pressures, both outside and inside of an RPC, along with the temperature, its leakage rate can be estimated. During the test period, the supporting button spacers inside an RPC may get detached due to manufacturing defect. This effect also needs to be detected.
}

\begin{document}

\maketitle

\section{Introduction}
The India-based Neutrino Observatory Project (INO) is a proposed underground high energy physics experiment at Theni, India to study the neutrino oscillation parameters using atmospheric neutrinos. The 50 kton magnetised INO-ICAL\cite{inowhite} detector will consist of 151 layers of iron altering with 150 layers of $\sim$2m\,$\times$\,2m Resistive Plate Chambers (RPC) as sensitive detectors. The ICAL detector will require approximately 30,000 RPCs and proposed to operate for 20 years. For success of the experiment, each of the RPCs has to function without showing any significant ageing during the period of operation. Hence, various tests are performed during and after production.
\begin{figure}[h!]
  \centering
  \includegraphics[width=0.6\textwidth]{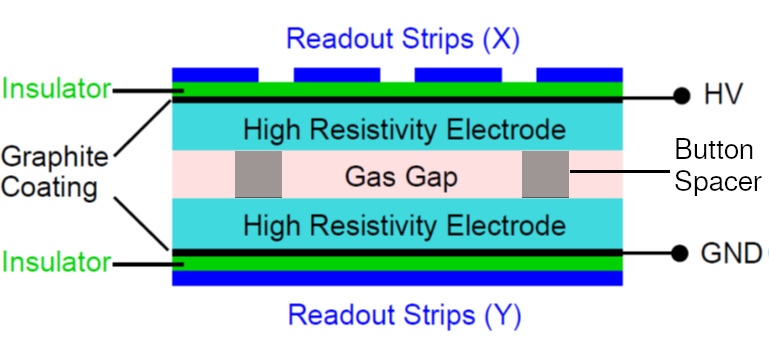}
  \caption{Schematic of a Resistive Plate Chamber.}
  \label{fig:rpc}
\end{figure}

The RPC is basically a gaseous based parallel-plate detector\cite{rpc_p1,rpc_p2}. The basic components of RPC are shown in Figure~\ref{fig:rpc}. The RPC is constructed using two parallel plates of glass having a bulk resistivity of the order of $10^{10}$-$10^{12}$\,$\Omega$\,cm with a gas gap of few mm. Uniform spacing between two glass plates is maintained using button spacers. Ideally, the whole chamber has to be leak proof. The outer surfaces of the glass plates are made conductive using graphite coating to form the electrodes where high voltages can be applied. The signal is readout by copper pickup panels on both sides of the RPC. In ICAL detector, the RPCs are going to be operated in avalanche mode with a gas mixture of R134a\,(95.2\%), iso-C$_4$H$_{10}$\,(4.5\%) and SF$_6$\,(0.3\%).\footnote{R134a is defined as 1,1,1,2-Tetrafluoroethane.} R134a gas acts as a target for the ionising particles passing through the gas gap. The iso-C$_4$H$_{10}$ absorbs the photons emitted in the recombination processes limiting the formation of secondary avalanches. SF$_6$, being an electronegative gas, localizes the signal in a small area to have better position resolution.

During the active operation of ICAL detector, more than 200,000\,litres of the gas mixtures will be circulating inside the 30,000 RPCs. To achieve this, a closed-loop gas circulation system (CLS) is designed whose main purpose is to recirculate the gas mixture, minimising wastage of gas which reduces the operational cost. The leakage of gas mixture from a closed-loop system will increase the cost of operation. Also, the leakage of outside atmosphere into the system will contaminate the gas mixture by water vapour and oxygen which may damage the RPC\cite{rpc_c,rpc_w}. The fluorine present in the RPC gas mixture will react with water vapour producing hydrofluoric acid which will damage the inner surfaces of the glass electrodes. The oxygen gas having affinity to electrons can affect the performance of the detector. Due to the aforementioned reasons, a proper leak test has to be performed on all the glass gaps at the time of production as well as during operation.
\begin{figure}
  \centering
  \includegraphics[width=0.6\textwidth]{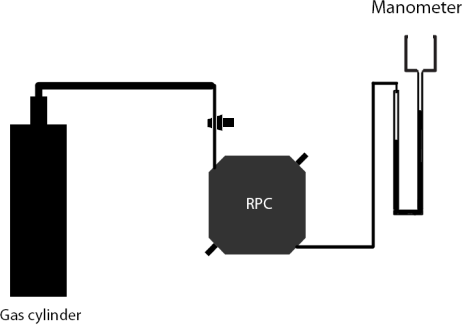}
  \caption{Schematics of a typical leak test setup.}
  \label{fig:test_rpc}
\end{figure}

The schematics of a typical leak test setup is shown in Figure~\ref{fig:test_rpc}. To estimate the leakage, the RPC is first pressurised up to 45\,mmWC above the atmospheric pressure and then sealed.\footnote{Millimetres water column, abbreviated to mmH$_2$O or mmWC, is a unit of pressure. It is the pressure required to support a water column of the specified height. 1\,mmWC\,$\simeq$\,0.098\,mbar. This will be the unit used in the paper.} Then the pressure inside the RPC is monitored over a long period of time using a conventional manometer. In a conventional manometer, the observable quantity is the differential pressure. The methods of leak rate calculation using conventional manometer are valid only when both the volume of the test subject and ambient pressure are kept constant. But both these quantities for a RPC gas gap depend widely on the ambient pressure and temperature which is constantly affected by the solar atmospheric tides and changes in weather.\footnote{The solar atmospheric tides are generated by the periodic heating of the atmosphere by the Sun. This regular diurnal cycle in heating generates tides in atmosphere that have periods related to the solar day.} Because of this effect, a large variation in the pressure is observed over a long period of time when measured from a conventional manometer. In such a scenario, if there is very small leak in the RPC, it is nearly impossible to detect. Also, it is not possible to get a proper quantitative estimation of the leak rate from such pressure measurements.

The principal aim of the current study is to test whether the RPC can hold the prescribed pressure difference or not in presence of the variation in ambient pressure due to solar atmospheric tides. The setup and technique discussed in this paper not only help to determine whether the RPC is leaking or not but also allows to estimate the quantity of the leakage. As the number of the RPCs to be used in ICAL detector is very large, the setup must be cost effective, time efficient and portable. The method described in this study is able to test multiple RPCs at the same time without moving them out of the storage area. The initial effort were made to achieve this goal and reported in reference \cite{Mondal_2016}. The methods and results discussed here are a continuation of that work.

As per the Poiseuille's law\cite{poiseuille}, the laminar flow rate of a fluid through a leak path is given in equation \ref{eq:poiseuille1}.
\begin{equation}
\left(\text{Flow Rate}\right)=\left(\text{Leak Constant}\right)\times\left(\text{Effective Pressure Difference}\right)\label{eq:poiseuille1}
\end{equation}
where, the \textit{Leak Constant} depends on the path of leakage (i.e. crack, hole, etc) and the viscosity of the gas mixture. The \textit{Leak Constant} quantifies the leakage in the system. The setup and techniques to calculate the \textit{Leak Constant}s for the gas gaps are discussed in this article.

\section{Experimental Setup}
\begin{figure}
  \centering
  \includegraphics[width=0.9\textwidth]{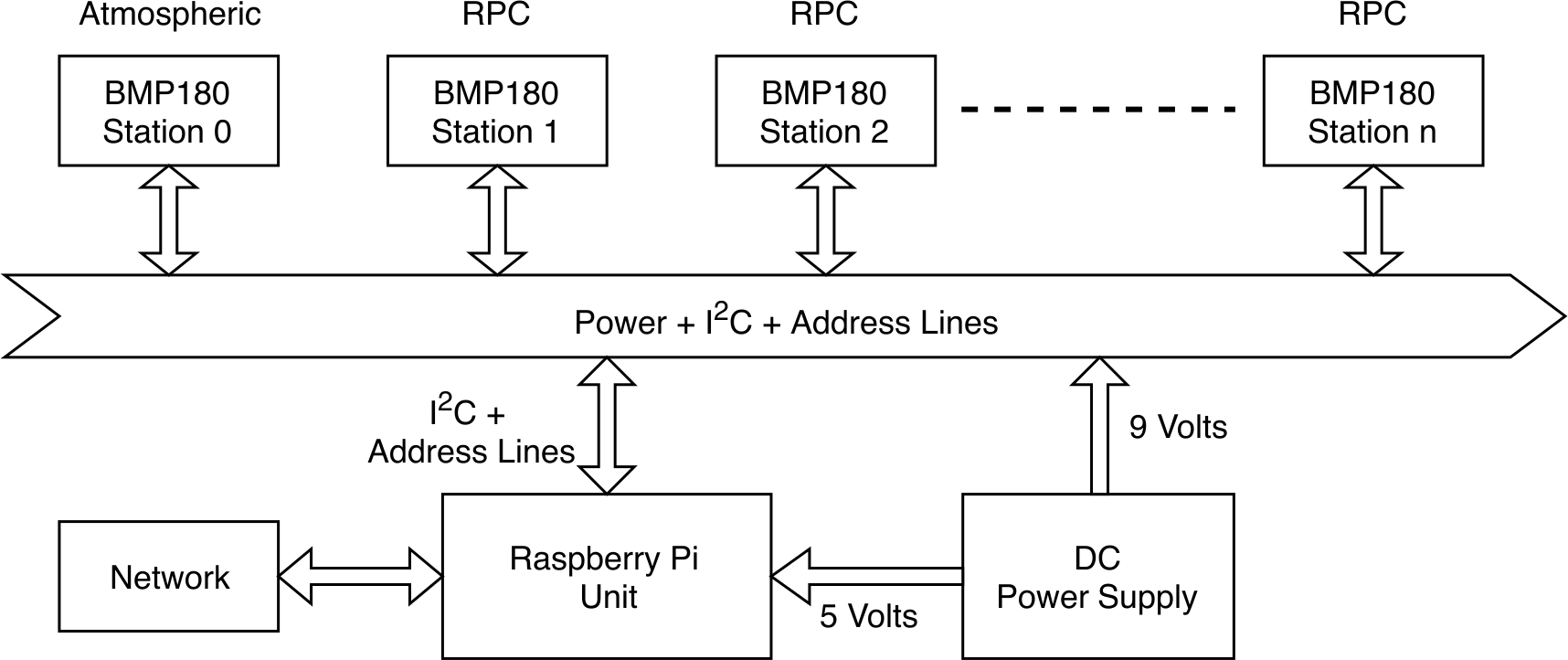}
  \caption{Schematics of the leak test setup.}
  \label{fig:schematics}
\end{figure}

\begin{figure}
  \centering
  \begin{subfigure}[b]{0.34\textwidth}
    \includegraphics[width=\textwidth]{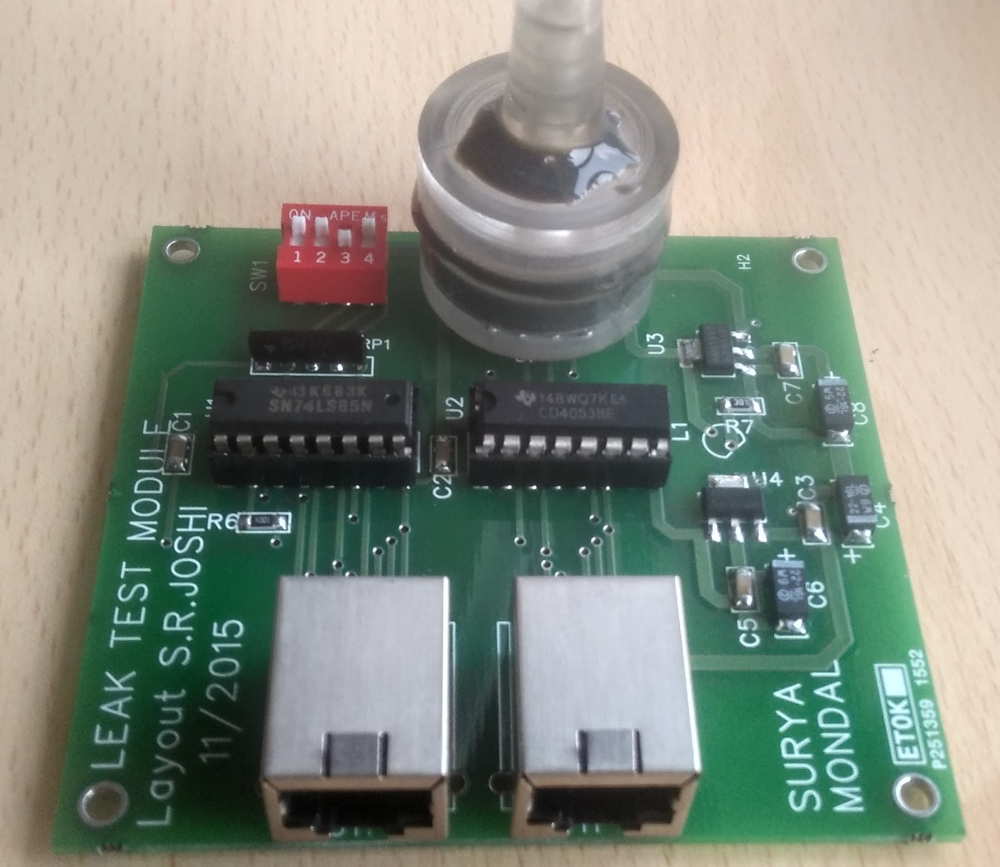}
    \caption{Leak Test Module.}
    \label{fig:pc3}
  \end{subfigure}
  \begin{subfigure}[b]{0.64\textwidth}
    \includegraphics[width=\textwidth]{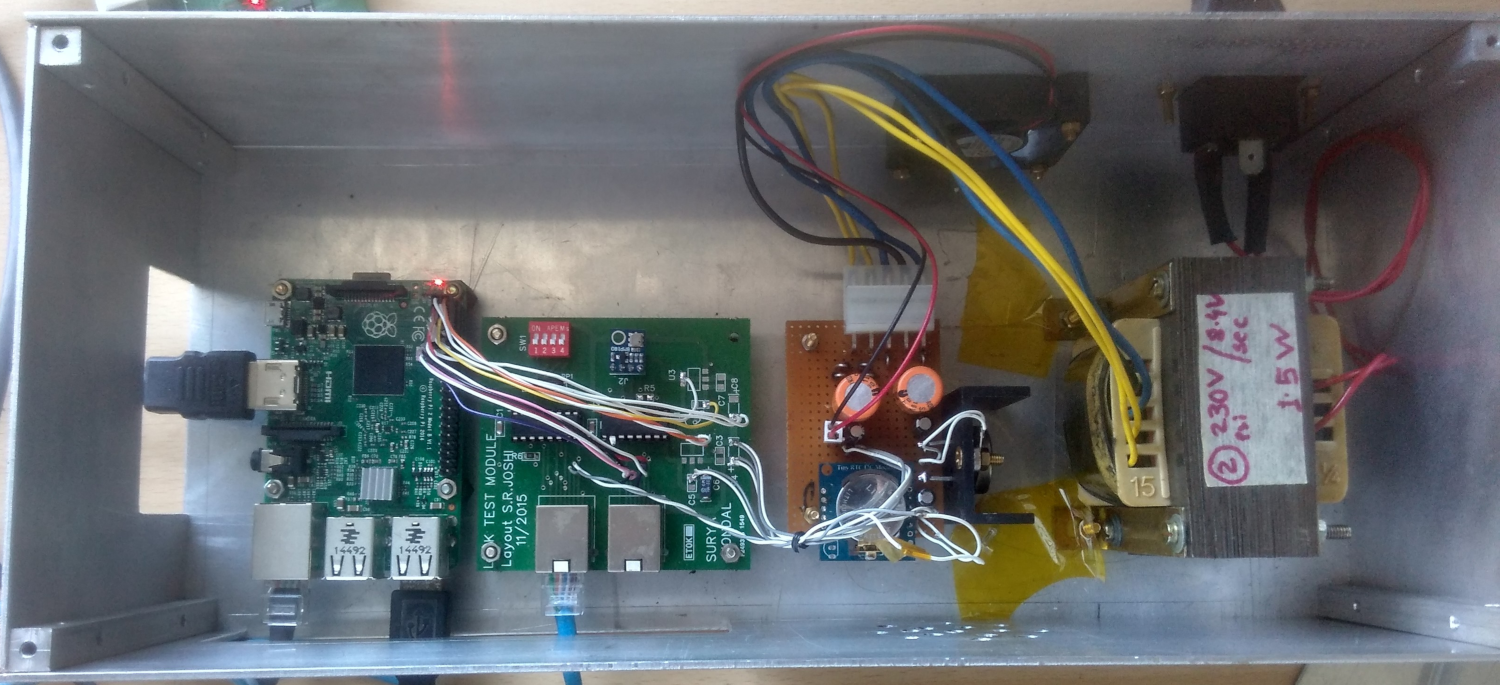}
    \caption{Raspberry Pi B \& Power Supply Module.}
    \label{fig:pc2}
  \end{subfigure}
  \caption{Leak Test Setup.}
  \label{fig:setup}
\end{figure}

The schematics of the experimental setup is shown in Figure~\ref{fig:schematics}. In this setup, instead of measuring the differential pressure using the conventional manometer, the absolute pressure and temperature inside the gas gap are measured using the sensor module, BMP180 manufactured by BOSCH\cite{bmp180}. The BMP180 is a piezo-resistive sensor having an accuracy of 0.7\,mmWC and 0.05\,$^{\circ}$C in the measurement of pressure and temperature respectively. This sensor is capable of record data samples for the minimum time interval of 76\,ms. The leak test module is shown in Figure~\ref{fig:setup}(\subref{fig:pc3}). One such module will record the pressure and temperature for one gas gap. The pressure and temperature data recorded by the module is readout using a \textit{Raspberry Pi\,v2\,B} (Pi) unit\cite{rpi} shown in Figure~\ref{fig:setup}(\subref{fig:pc2}). The data is stored on the on-board memory of the Pi unit. As shown in Figure~\ref{fig:setup}(\subref{fig:pc3}), each module has two bus ports. This allows to daisy-chain several leak test modules and can be controlled from a single Pi unit.

The common bus mainly consists of Power, Data and Address lines. To avoid the voltage drop in the supply line over long distance, 9\,V DC is supplied from the Pi End and converted to required voltage at each test module. A 4-bit DIP switch is used on each module to set a unique address for itself. The Pi acquires the data from each station by selecting its unique address. So, only one test module is allowed to communicate with the control unit at a time. As 4-bit address lines are used in this setup, a maximum of fifteen gas gaps can be tested simultaneously. The system is scalable to handle more gas gaps at the same time, by simply adding more address lines. One of the leak test module is dedicated to record the ambient pressure and temperature for the test duration.

The data from the BMP180s can also be acquired without wires by microcontrollers equipped with WiFi modules (i.e. NodeMCU module\cite{nodemcu2015}) which utilise the WiFi network at the premises to send the data to a computer on the same network eliminating the need of the wired common bus and Pi Unit.

In the current setup, the pressure and temperature data from each sensor is recorded continuously with a specified interval of $3$\,seconds. The final data recorded for a gas gap has the ambient pressure and temperature, the gas gap pressure and temperature and the time stamp for each measurement. Using these measurements, the method to quantify the leakage is discussed in the Section \ref{sec:calculation}.

\section{Detection of Button Pop-Ups}\label{sec:button}
\begin{figure}
  \centering
  \includegraphics[width=0.99\textwidth]{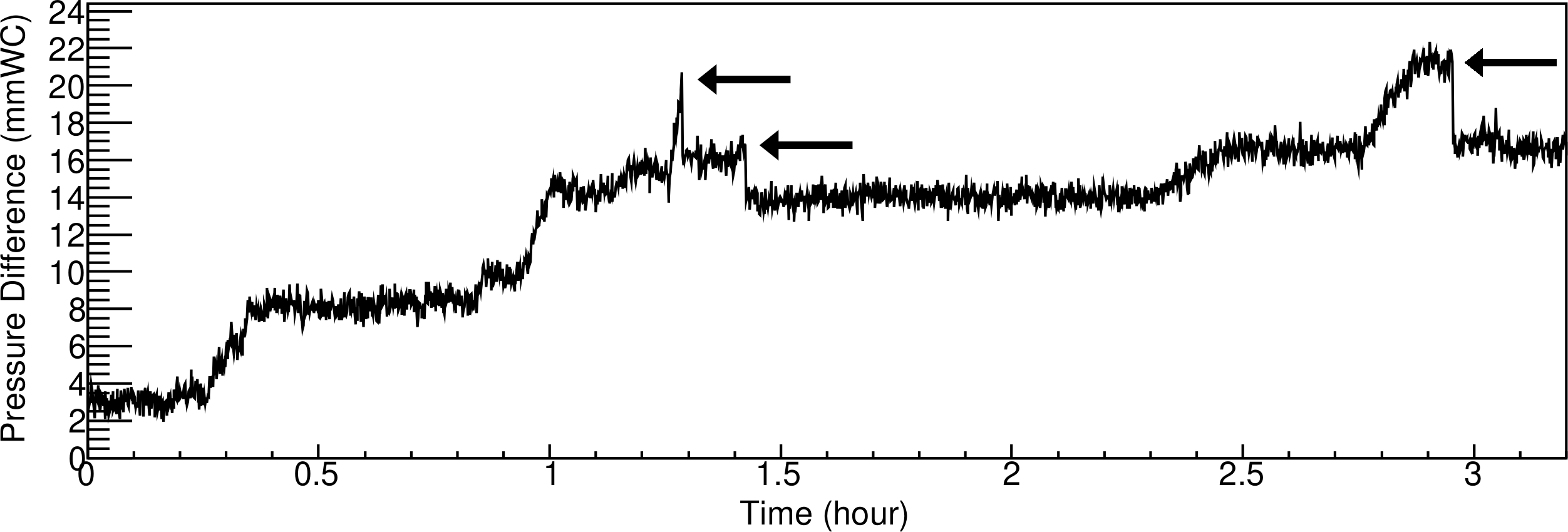}
  \caption{Variation of pressure difference with time showing button pop-ups.}
  \label{fig:button}
\end{figure}

The structural stability of the RPC is maintained by the buttons. However, it is observed that sometimes the glue (used to attach the buttons to the glass plates) fails to hold under pressure which results in detaching of the buttons from the glass plates.\footnote{Hereafter such an event is referred to as `button pop-up'.} This will weaken the structure of RPC. Also during detector operations, this will increase the spacing between the glass plates, thus decreasing the effective electric field resulting in reduced signal strength. Also, for any working RPC, it is necessary that the glue used to attach the buttons to the glass plates continues to hold under pressure. For any RPC gas gap, even one button is not attached, that glass gap will not be suitable to hold more pressure as eventually the glue for more and more buttons will give way making the gas gap weaker. Hence, it is essential to detect any `button pop-up' events during the leak test.

With each `button pop-up' event, the volume of RPC gap increases which in turn results in a decrease in the pressure inside the gap. Thus the pressure difference between the outside and inside of the gap decreases. In the plot of pressure difference with time, this effect will be observed as a sudden drop in the pressure difference. Figure~\ref{fig:button} shows the variation of pressure difference with time for a RPC where there are `button pop-up' events. It can be observed in the figure that there are three `button pop-up' events (pointed by arrows) in this RPC. As seen here, these `button pop-up' events results in decrease of the pressure difference in a fraction of time. These events cannot be detected with conventional manometer unless they pressure is recorded continuously using a precise differential pressure sensor. Hence, the apparatus, described in the current paper, is very helpful in detecting the `button pop-up' events during the test.

\section{Leak Rate Calculation}\label{sec:calculation}
\begin{figure}[h!]
  \centering
  \includegraphics[width=0.99\textwidth]{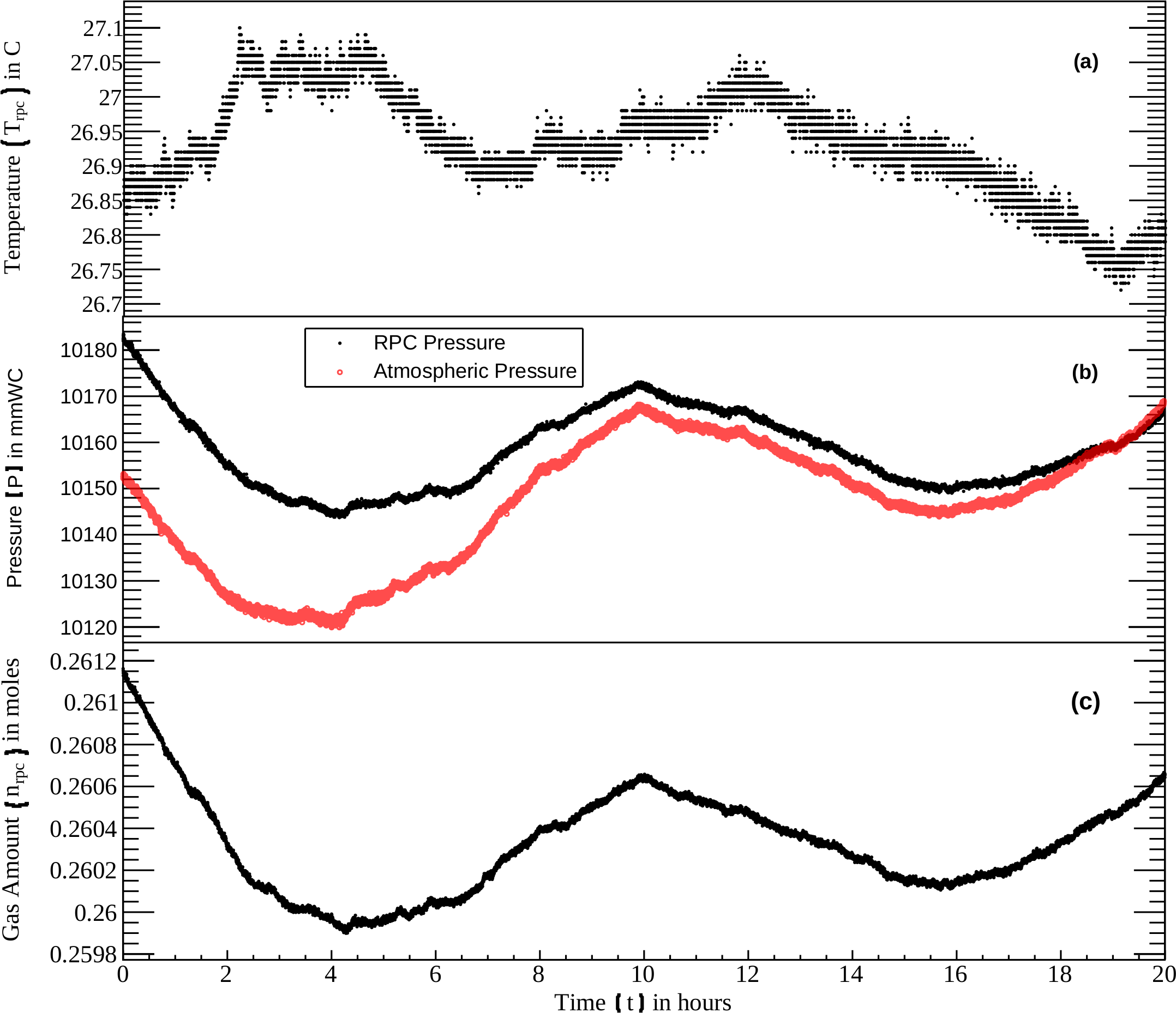}
  \caption{\textbf{(a)} Variation of temperature ($T_{\textrm{rpc}}$) with time, \textbf{(b)} Variation of atmospheric ($P_{\textrm{atm}}$ : Black) and RPC ($P_{\textrm{rpc}}$ : Red) pressure with time, \textbf{(c)} Amount of gas in the RPC with time assuming $V_{\textrm{rpc}}=6.3$\,litres for sample RPC gap-1.}
  \label{fig:temp}
\end{figure}

The variation of temperature of the gas gap and variation of pressure in the gas gap as well as atmosphere respectively with time are illustrated in Figure \ref{fig:temp}(a) and \ref{fig:temp}(b). It can be observed that the pressure inside the gap is following the trend of atmospheric pressure. This implies that the volume of the gap changes with the change of atmospheric pressure. Now, from the Ideal Gas Law, the amount of gas ($n$), inside a chamber of volume $V$, can be calculated at time $t$ using equation equation \ref{mole}.\footnote{All the values of $P$, $V$ and $T$ are converted suitably to calculate the value of $n$ in mole. Suffix `rpc' denotes the measurements acquired from inside the RPC.}
\begin{equation}
  n_{\textrm{rpc}\textbar t}=\frac{P_{\textrm{rpc}\textbar t}V_{\textrm{rpc}\textbar t}}{RT_{\textrm{rpc}\textbar t}} \label{mole}
\end{equation}\\
where, $R$ is the ideal gas constant having value 8.314\,J\,mole$^{-1}$\,K$^{-1}$. From the dimensions of the gas gap, the volume of the RPC is estimated to approximately 6.3\,litres. Taking this into consideration, the amount of gas inside the gap can be calculated using equation \ref{mole}. Figure~\ref{fig:temp}(c) shows the variation of the amount of gas within the gap with respect to time. The instantaneous leak rate of a gap can be quantified as the slope of this plot.

\begin{figure}
  \centering
  \includegraphics[width=0.99\textwidth]{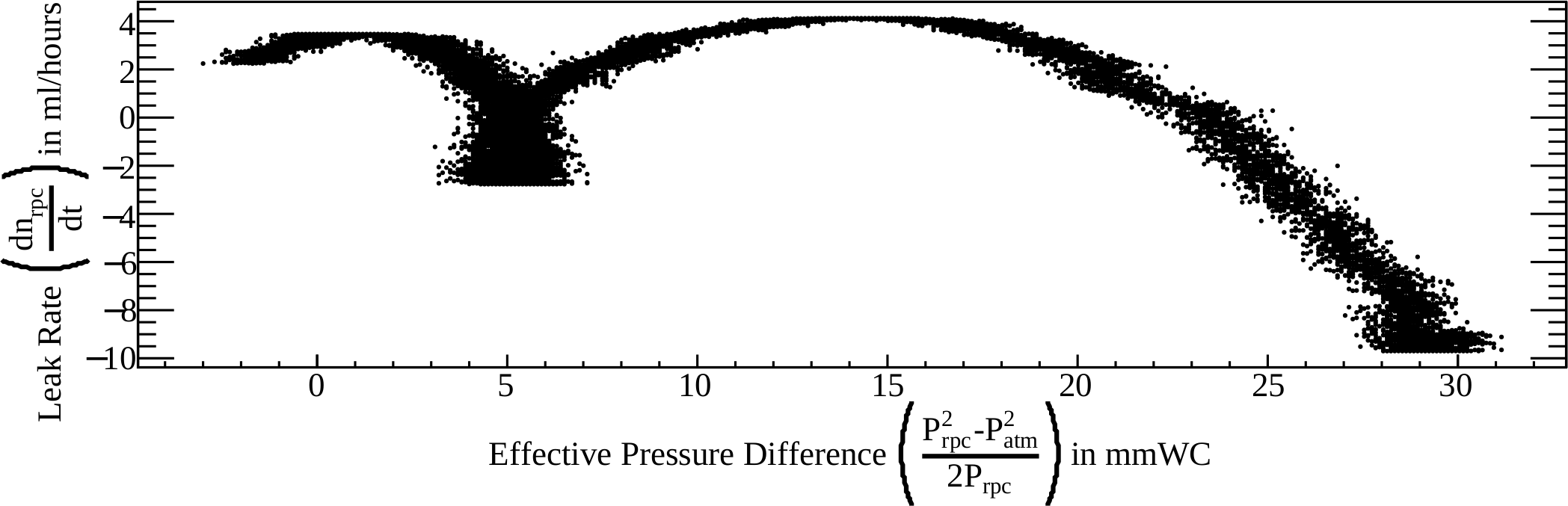}
  \caption{$\left(\frac{\mathrm{d}n_{\textrm{rpc}}}{\mathrm{d}t}\right)$ vs $\frac{P_{\textrm{rpc}}^{2}-P_{\textrm{atm}}^{2}}{2P_{\textrm{rpc}}}$ in sample RPC gap-1 without correction.}
  \label{fig:preQt}
\end{figure}

To estimate the absolute leak rate, the Poiseuille's equation for compressible fluids\cite{poiseuille} is used. The Poiseuille's equation of Leak Rate $\left(\frac{\mathrm{d}n_{\textrm{rpc}}}{\mathrm{d}t}\right)$ for compressible fluids is given in equation \ref{eq:poiseuille},
\begin{equation}
  \left.\frac{\mathrm{d}n_{\textrm{rpc}}}{\mathrm{d}t}\right| _t=\textrm{C}_{\textrm{Leak}}\times\left(\frac {P_{{\textrm{rpc}\textbar t} }^{2}-P_{{\textrm{atm}\textbar t} }^{2}}{2P_{{\textrm{rpc}\textbar t} }}\right)\label{eq:poiseuille}
\end{equation}
where, $\textrm{C}_{\textrm{Leak}}$ depends on the path of leakage (i.e. crack, hole, etc) and the viscosity of the gas mixture and it quantifies the leakage in the system.\footnote{The viscosity of the gas is assumed to be constant over the small changes of room temperature during the test period. In case of large changes in temperature, the changes in the value of viscosity are also needed to be considered.} Figure~\ref{fig:preQt} shows the leak rate $\left(\frac{\mathrm{d}n_{\textrm{rpc}}}{\mathrm{d}t}\right)$ which is calculated from Figure~\ref{fig:temp}(c) as a function of the effective pressure difference $\left(\frac{P_{\textrm{rpc}}^{2}-P_{\textrm{atm}}^{2}}{2P_{\textrm{rpc}}}\right)$.\footnote{It is assumed that mole\,$\simeq$\,22.4\,litres for better visualisation.} According to equation \ref{eq:poiseuille}, it is expected to behave as a straight line passing through the origin with the slope quantifying the leakage in the system but it can be clearly observed that it is not a straight line.

Also, the gas gap under test is sealed from the inlet. Hence, no more gas is getting filled but the Figure~\ref{fig:temp}(c) shows an apparent increase in the amount of gas inside the gap. This effect can be explained with the change in the volume of the gas gap due to the variations in the atmospheric pressure and the room temperature. The change in the volume of the gas gap, is unknown and cannot be measured. To compensate this change in volume, the volume of the RPC gap at time $t$ is represented by the equation \ref{vcterm}.
\begin{equation}
  V_{\textrm{rpc}\textbar t} = V_{\textrm{rpc}}\left(1-x_T\left(T_{\textrm{rpc}\textbar t}-T_{\textrm{rpc}\textbar t=0}\right)\right)\left(1-x_P\left(P_{\textrm{atm}\textbar t}-P_{\textrm{atm}\textbar t=0}\right)\right)\label{vcterm}
\end{equation}
where, $T_{\textrm{rpc}\textbar t=0}$ and $P_{\textrm{atm}\textbar t=0}$ are equal to $T_{\textrm{rpc}}$ and $P_{\textrm{atm}}$ at time\,$t=0$, respectively.\footnote{Suffix `atm' denotes the measurements acquired from atmosphere.} $V_{\textrm{rpc}}$ is 6.3\,litres. Assuming that the change in volume is linear to both the atmospheric pressure and the room temperature, two independent linear correction terms ($x_P$ and $x_T$) are introduced.\footnote{Present method of estimation of leak cannot handle change in volume caused by `button pop-up' event during the period of leak test.} Using the volume calculated in equation \ref{vcterm} in the equation \ref{mole}, $n_{\textrm{rpc}}$ is represented in equation \ref{ct}.
\begin{equation}
  n_{\textrm{rpc}\textbar t}=\left(\frac{V_{\textrm{rpc}}}{R}\right)\left(\frac{P_{\textrm{rpc}\textbar t}}{T_{\textrm{rpc}\textbar t}}\right)\left(1-x_T\left(T_{\textrm{rpc}\textbar t}-T_{\textrm{rpc}\textbar t=0}\right)\right)\left(1-x_P\left(P_{\textrm{atm}\textbar t}-P_{\textrm{atm}\textbar t=0}\right)\right) \label{ct}
\end{equation}
\begin{figure}
  \centering
  \includegraphics[width=0.8\textwidth]{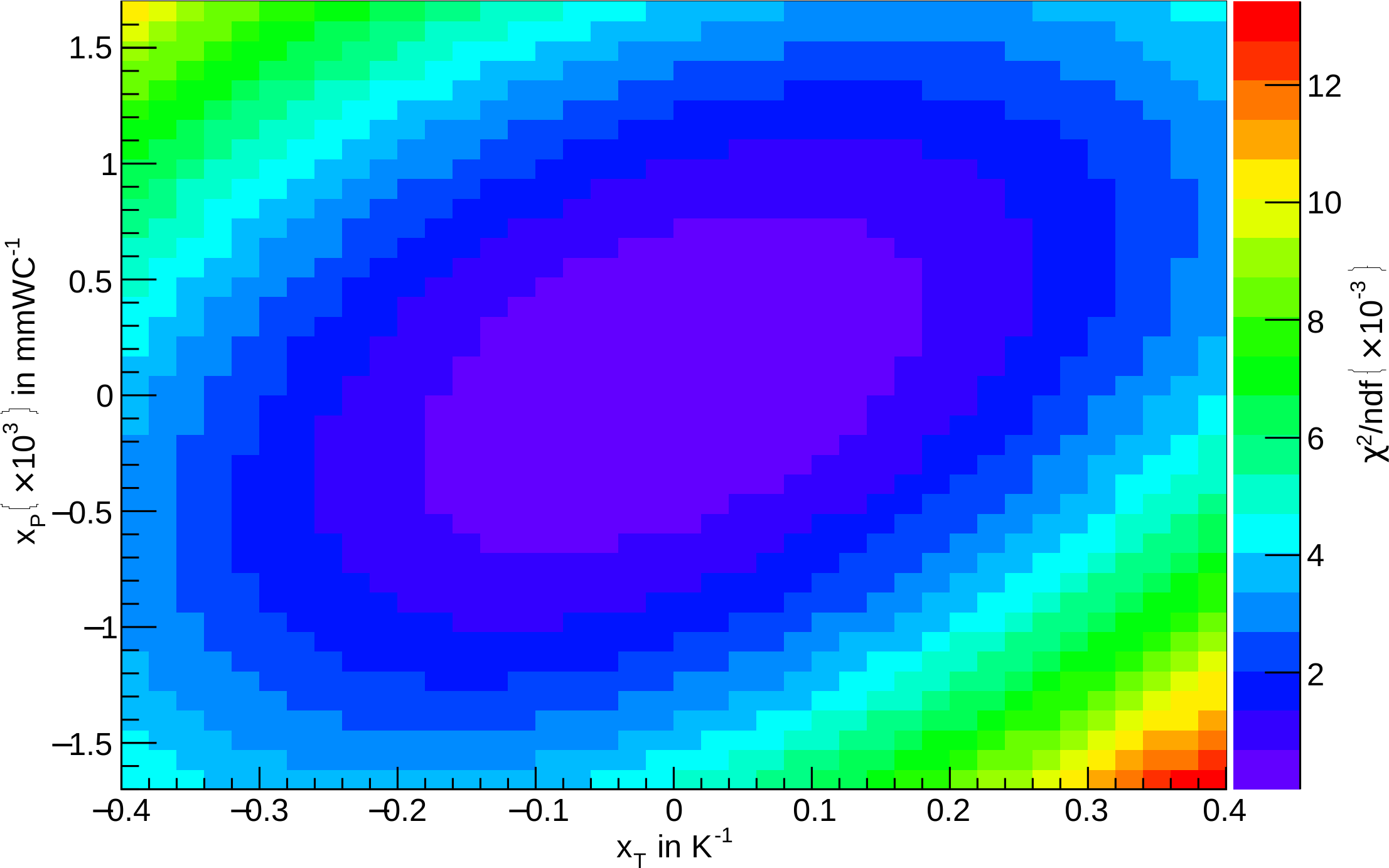}
  \caption{$\chi^2/\textrm{ndf}$ values of the straight line fit for the plots of $\frac{\mathrm{d}n_{\textrm{rpc}}}{\mathrm{d}t}$ vs $\frac{P_{\textrm{rpc}}^{2}-P_{\textrm{atm}}^{2}}{2P_{\textrm{rpc}}}$ for different combinations of $x_T$ and $x_P$ for sample RPC gap-1.}
  \label{fig:xp}
\end{figure}
\begin{figure}
  \centering
  \includegraphics[width=0.99\textwidth]{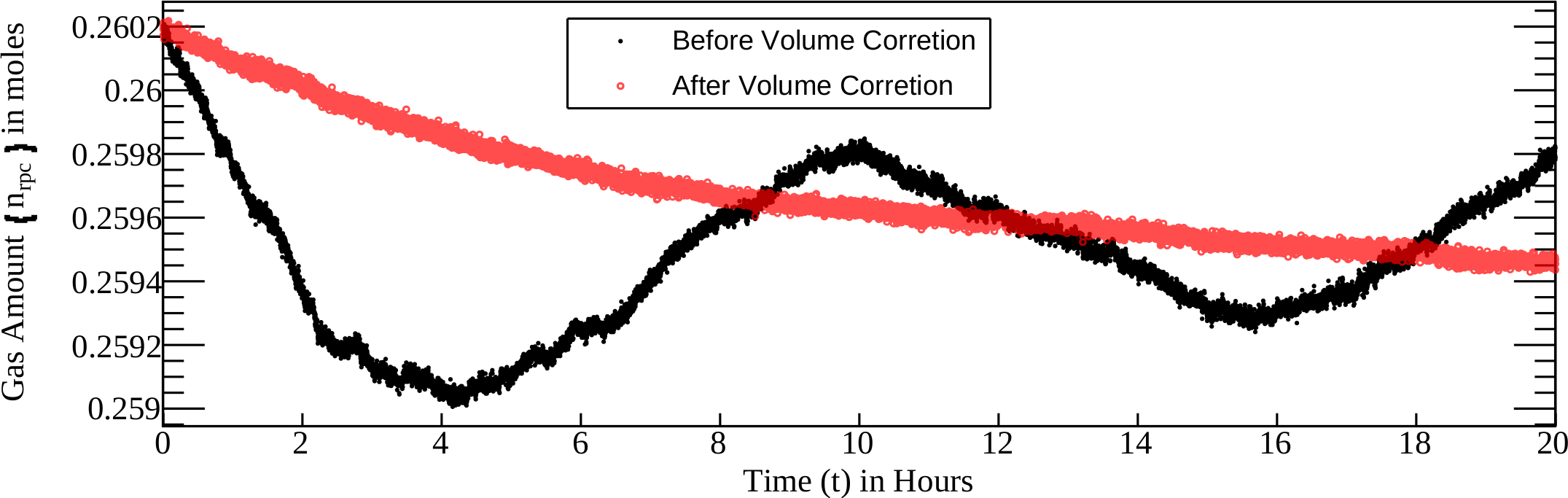}
  \caption{Amount of gas in the RPC before (Black) and after (Red) correction for sample RPC gap-1.
  }
  \label{fig:with}
\end{figure} 

For different values of $x_T$ and $x_P$, $n_{\textrm{rpc}}$ is calculated using equation \ref{ct} and then is plotted against time as shown in Figure~\ref{fig:temp}(c). For each combination of the correction terms, the plot of $n_{\textrm{rpc}}$ versus time is used to calculate $\frac{\mathrm{d}n_{\textrm{rpc}}}{\mathrm{d}t}$.
The plot of $\frac{\mathrm{d}n_{\textrm{rpc}}}{\mathrm{d}t}$ vs $\frac{P_{\textrm{rpc}}^{2}-P_{\textrm{atm}}^{2}}{2P_{\textrm{rpc}}}$ is then fitted with a straight line and $\chi^2/\textrm{ndf}$ is calculated. The $\chi^2/\textrm{ndf}$ values for each combination of $x_T$ and $x_P$ are shown in Figure~\ref{fig:xp}. For particular combination of $x_T$ and $x_P$, the $\chi^2/\textrm{ndf}$ will be minimum which is shown in Figure~\ref{fig:xp}. To minimise the uncertainties at the minimum $\chi^2/\textrm{ndf}$, the procedure is repeated for multiple iterations with subsequently smaller range of $x_T$ and $x_P$ to obtain the final values. In the current case of sample RPC gap-1, the correction terms at minimum $\chi^2/\textrm{ndf}$ are found to be 
\[x_T=-3.23\times10^{-3}\textrm{\,K}^{-1}\textrm{ and }x_P=7.83\times10^{-5}\textrm{\,mmWC}^{-1}\textrm{.}\]

\begin{figure}
  \centering
  \includegraphics[width=0.99\textwidth]{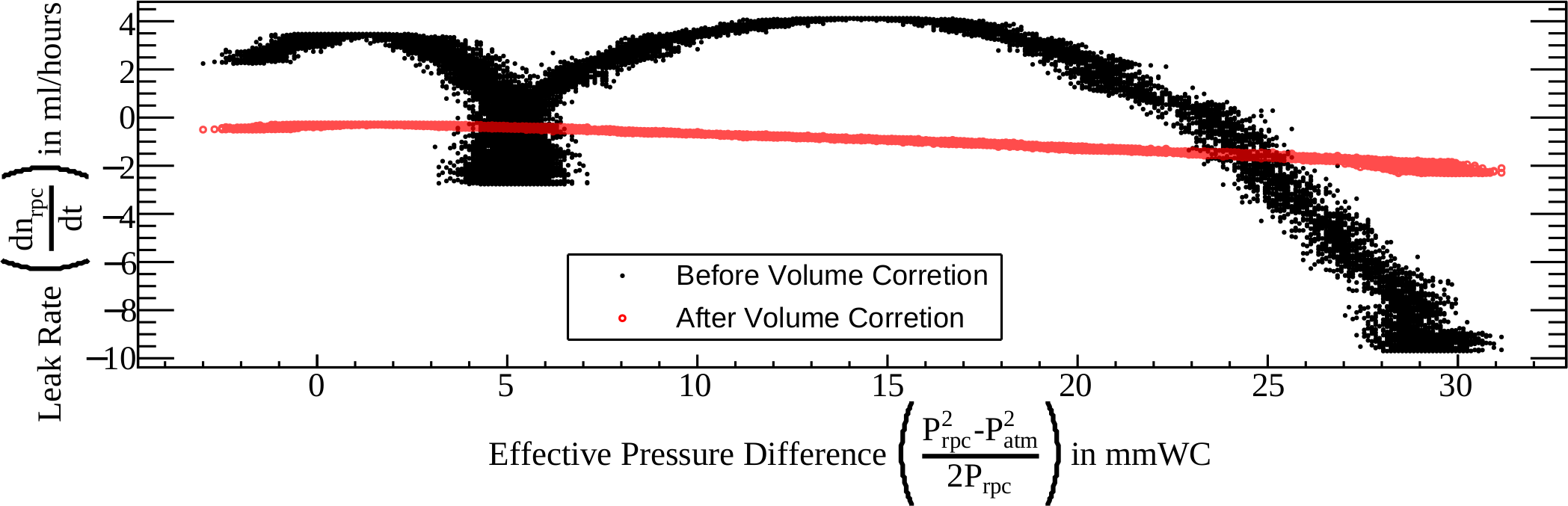}
  \caption{$\frac{\mathrm{d}n_{\textrm{rpc}}}{\mathrm{d}t}$ vs $\frac{P_{\textrm{rpc}}^{2}-P_{\textrm{atm}}^{2}}{2P_{\textrm{rpc}}}$ plots before (Black) and after (Red) correction for sample RPC gap-1.}
  \label{fig:qt}
\end{figure}

The negative value of $x_T$ implies that the volume of the RPC gap increases with increase in room temperature an the positive value of $x_P$ means that the volume of RPC gap decreases with increase in atmospheric pressure. The variation of amount of gas inside the RPC gap, $n_{\textrm{rpc}}$ which is calculated with these correction factors with time is presented in Figure~\ref{fig:with}. Comparing this with the variation observed before this volume correction, the increase in the amount of gas inside the gap, is not observed which implies that the proper volume correction is applied here. The leak rate versus the effective pressure difference is shown in Figure~\ref{fig:qt} which shows a nice straight line behaviour as expected from the Poiseuille's equation. The value of $\textrm{C}_{\textrm{Leak}}$ calculated from this is
\[\textrm{C}_{\textrm{Leak}}=-\left(6.73\pm 0.007\left(\textrm{stat}\right)\right)\times 10^{-2}\textrm{\,ml\,hour$^{-1}$\,mmWC$^{-1}$}.\]

Here, the negative value of $\textrm{C}_{\textrm{Leak}}$ indicates that the leak is from inside to outside. It means that if a constant pressure difference of 20\,mmWC is maintained between outside and inside of this RPC gap, then it will leak 1.442\,$m$mol or 32.3\,ml of gas within 24\,hours. Here, the value of $\textrm{C}_{\textrm{Leak}}$ clearly shows that this gap is having a leak.

\begin{figure}
  \centering
  \includegraphics[width=0.99\textwidth]{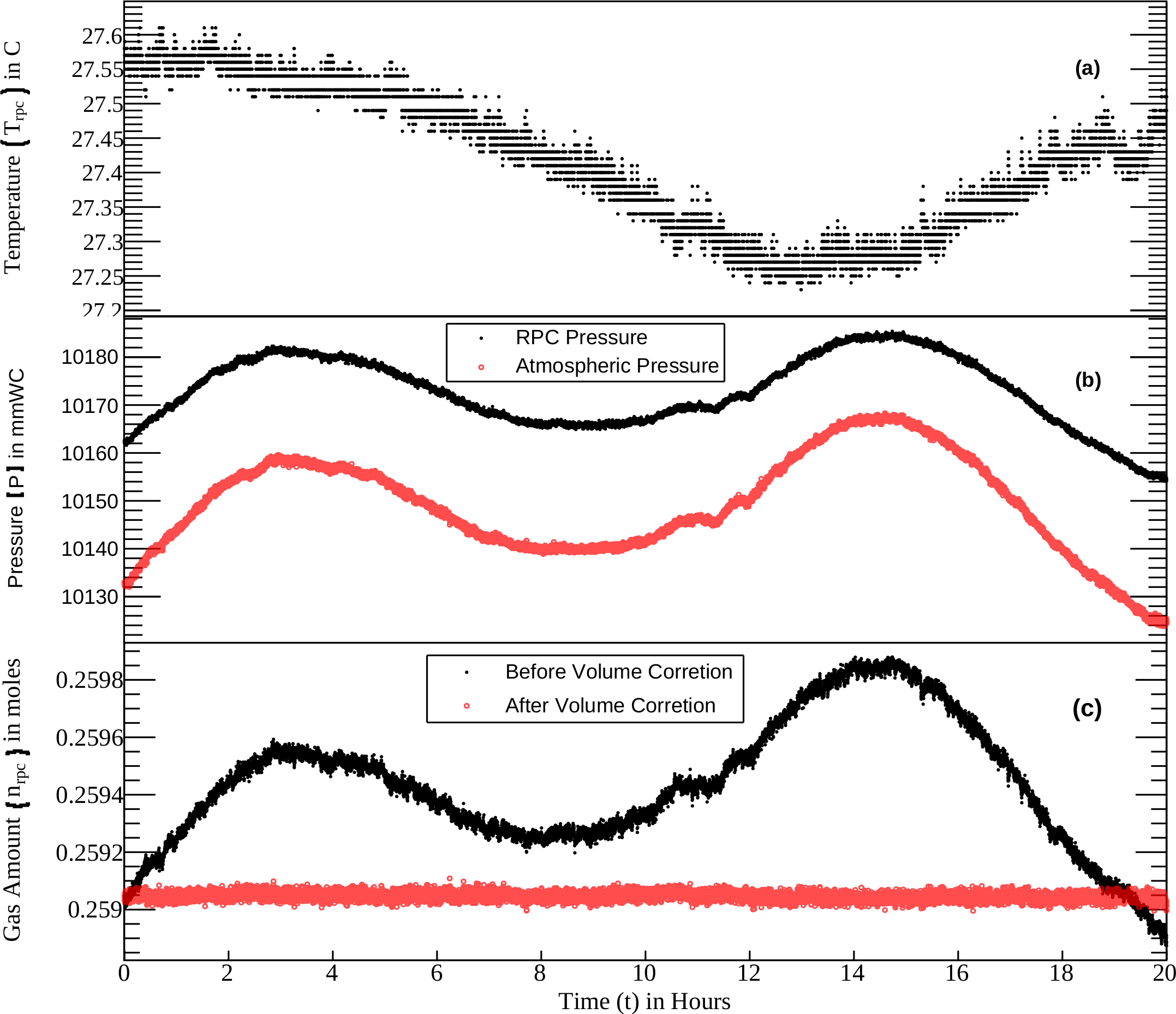}
  \caption{\textbf{(a)} Variation of temperature ($T_{\textrm{rpc}}$) with time, \textbf{(b)} Variation of atmospheric ($P_{\textrm{atm}}$ : Black) and RPC ($P_{\textrm{rpc}}$ : Red) pressure with time, \textbf{(c)} Amount of gas in the RPC before (Black) and after (Red) correction for sample RPC gap-2.}
  \label{fig:with1}
\end{figure}

\begin{figure}
  \centering
  \includegraphics[width=0.99\textwidth]{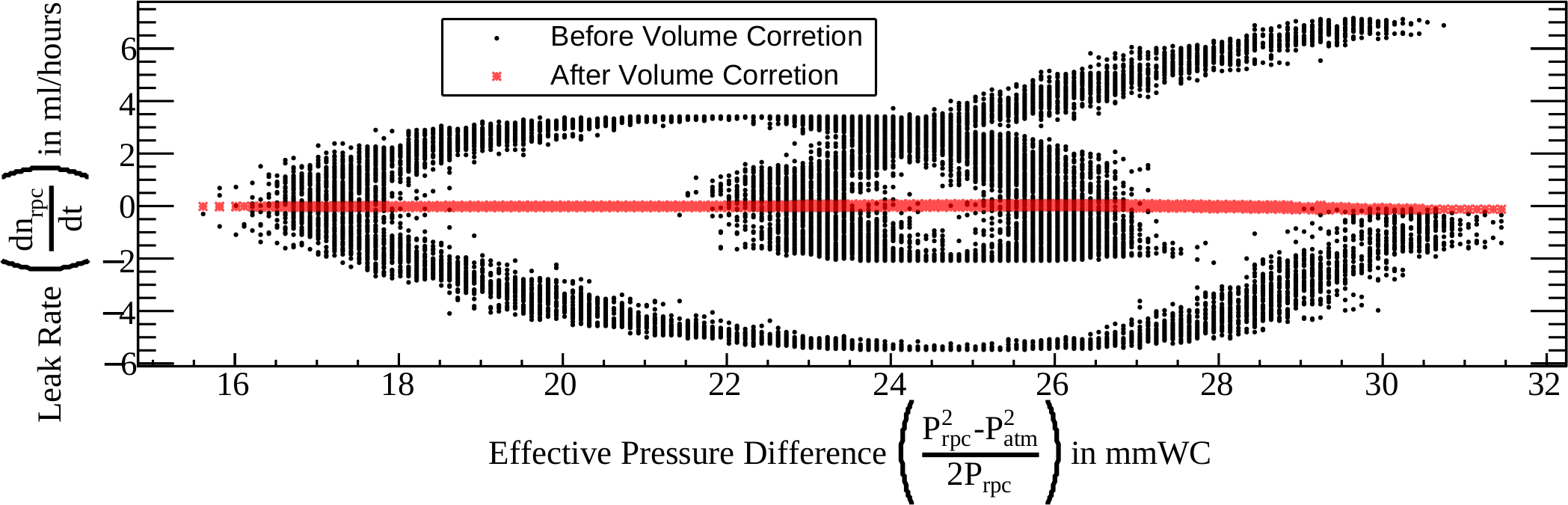}
  \caption{$\frac{\mathrm{d}n_{\textrm{rpc}}}{\mathrm{d}t}$ vs $\frac{P_{\textrm{rpc}}^{2}-P_{\textrm{atm}}^{2}}{2P_{\textrm{rpc}}}$ plots before (Black) and after (Red) correction for sample RPC gap-2.}
  \label{fig:qt1}
\end{figure}

Figure~\ref{fig:with1} and \ref{fig:qt1} show the test results for another RPC gap (namely, gap-2). From Figure~\ref{fig:qt1}, the value of $\textrm{C}_{\textrm{Leak}}$ for this RPC gap is estimated to be
\[\textrm{C}_{\textrm{Leak}}=-\left(5.1\pm 0.15\left(\textrm{stat}\right)\right)\times 10^{-4}\textrm{\,ml\,hour$^{-1}$\,mmWC$^{-1}$}.\]
In this case, if a constant pressure difference of 20\,mmWC is maintained between outside and inside of this RPC gap, then it will leak 10.93\,$\mu$mol or 0.245\,ml of gas within 24\,hours. From the small value of $\textrm{C}_{\textrm{Leak}}$, it can be observed that this gap has comparably very small leak.

\begin{figure}
  \centering
  \includegraphics[width=0.99\textwidth]{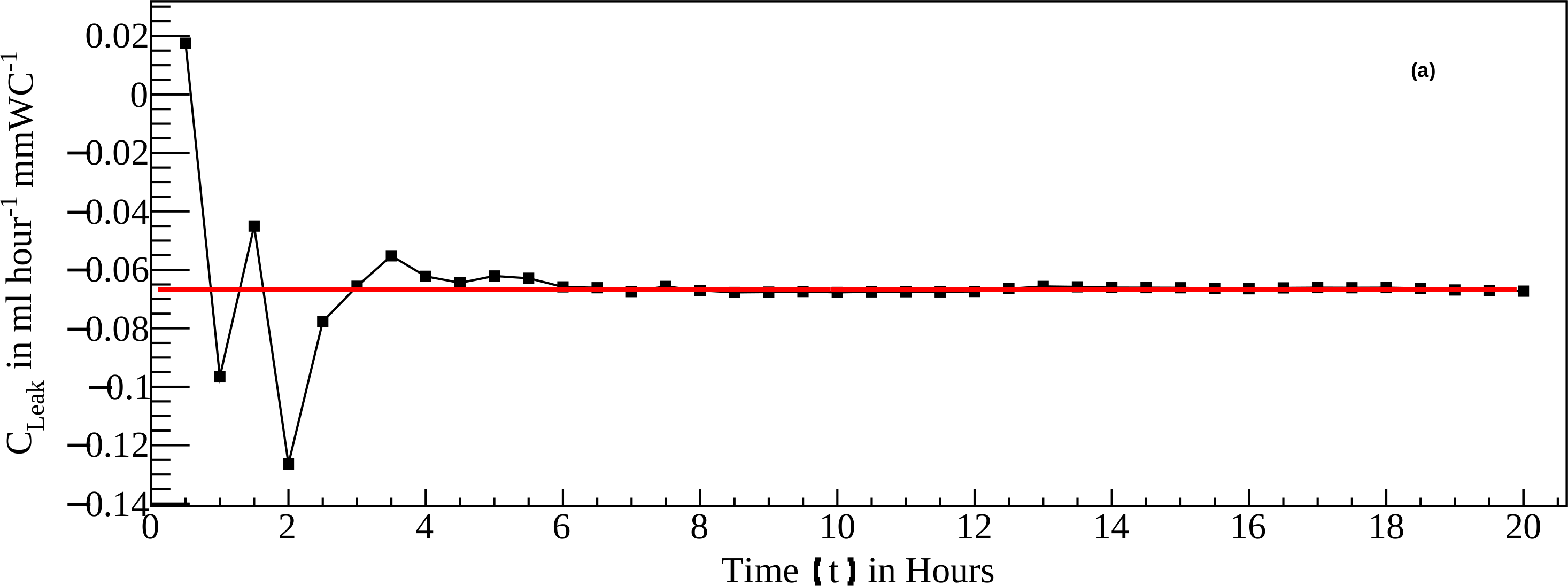}
  \vspace*{10pt}
  \includegraphics[width=0.99\textwidth]{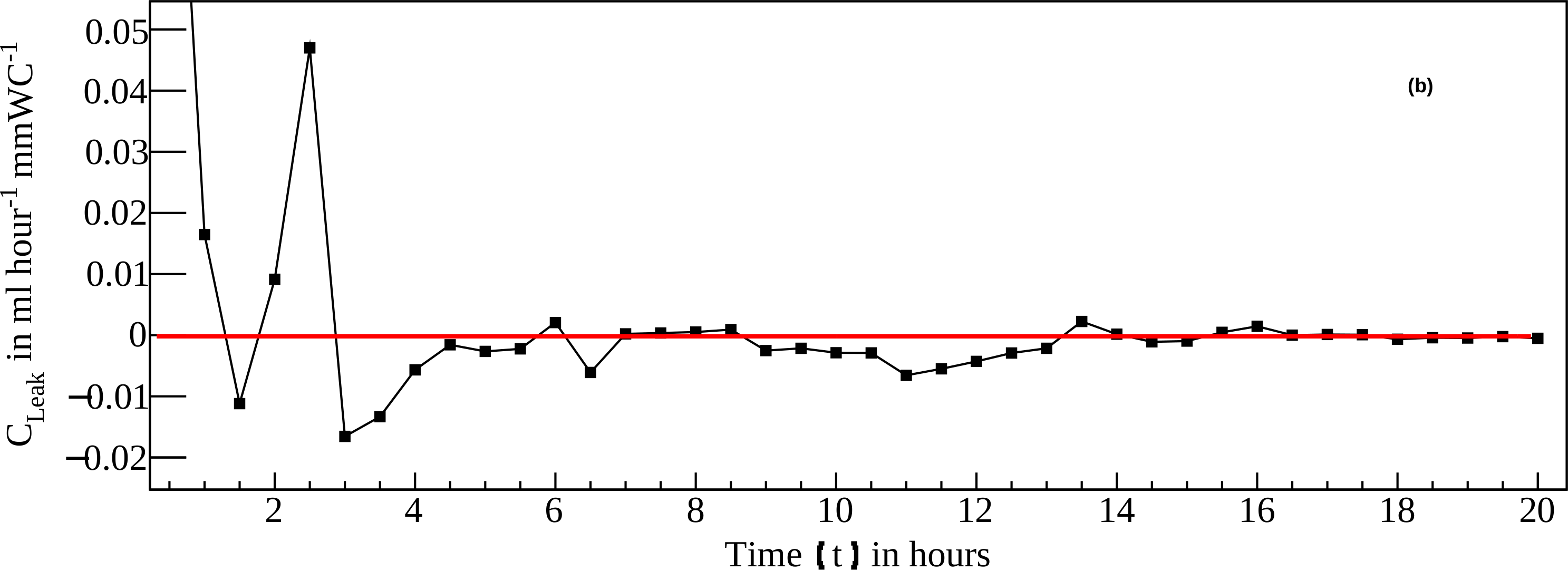}
  \caption{Leak rate estimation for data sets of different length for (a) RPC gap-1 and (b) RPC gap-2.}
  \label{fig:time}
\end{figure}

As seen from Figure~\ref{fig:temp}, the RPC gap was under the test for a duration of about 20 hours. One of the aim of the current study is to estimate the optimum/minimum time required to calculate the leak rate satisfactorily. Thus, the same study was performed by splitting the data into various sets for different duration from the start time of leak test and the value of $\textrm{C}_{\textrm{Leak}}$ is calculated. In Figure~\ref{fig:time}, the value of $\textrm{C}_{\textrm{Leak}}$ for different duration of time is plotted. For RPC gap-1, it can be observed that a minimum of 7-8 hours is required to estimate the leakage. This method requires a significant amount of data to fit the $\frac{\mathrm{d}n_{\textrm{rpc}}}{\mathrm{d}t}$ vs $\frac{P_{\textrm{rpc}}^{2}-P_{\textrm{atm}}^{2}}{2P_{\textrm{rpc}}}$ plots in order to get proper results. It should be noted that the minimal time required for a test will depend on the quantity of the leakage and also the environmental conditions during the test. If the leak rate is very small then more time is needed and vice versa. In the case of RPC gap-2, it can be observed from Figure~\ref{fig:time}(b) that about 7 hours is required to estimate leak rate with an uncertainty of $\sim 2\times 10^{-3}$\,ml\,hour$^{-1}$\,mmWC$^{-1}$, but about 15 hours is needed to have a result with uncertainty less than $\sim 2\times 10^{-4}$\,ml\,hour$^{-1}$\,mmWC$^{-1}$.

It would also be noted that the calculation of $\textrm{C}_{\textrm{Leak}}$ is directly dependent on the value of $V_{\textrm{rpc}}$. The relative error in measurement of $V_{\textrm{rpc}}$ will directly propagate to the relative error in calculation of $\textrm{C}_{\textrm{Leak}}$.

\section{Systematic Error}
\begin{figure}
  \vspace*{-15pt}
  \centering
  \includegraphics[width=0.99\textwidth]{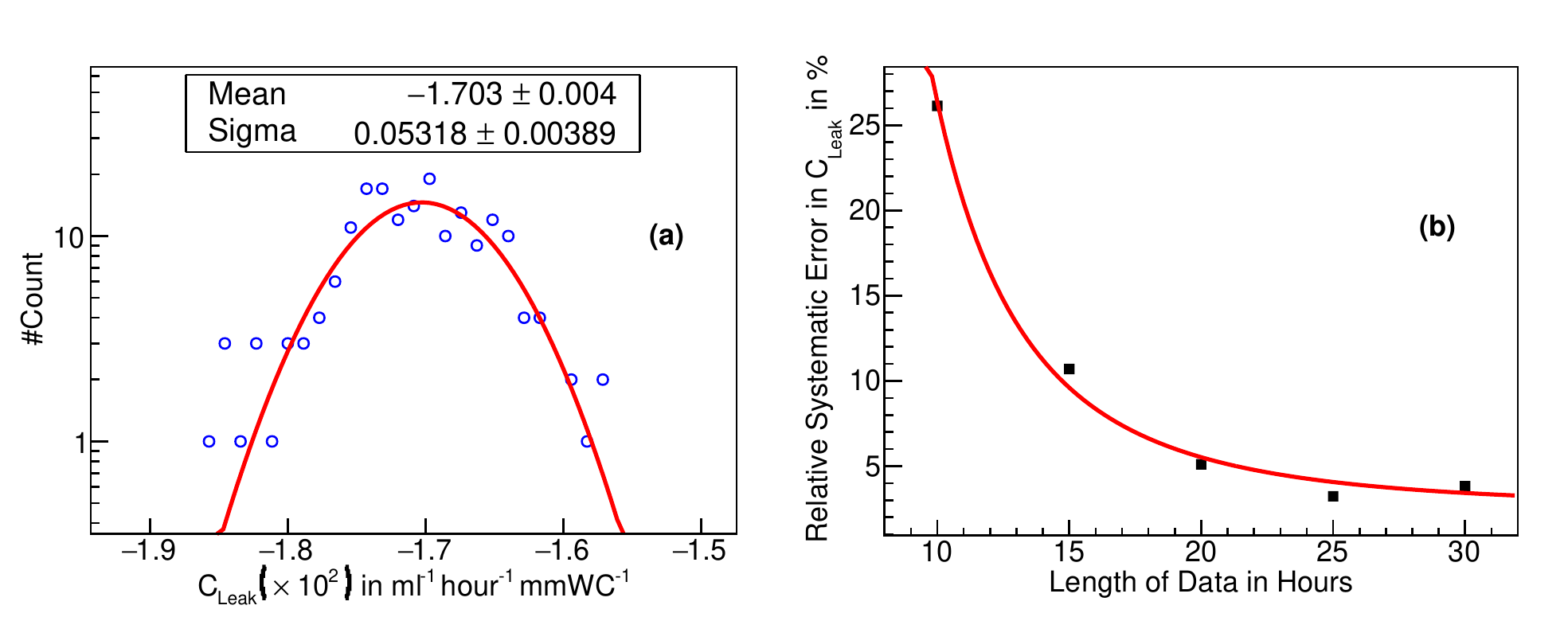}
  \caption{\textbf{(a)} Systematic error with data sets with length of 25 hours, \textbf{(b)} Variation of relative systematic error with length of data.}
  \label{fig:systematic}
\end{figure}

To estimate the systematic error in the estimation of $\textrm{C}_{\textrm{Leak}}$, one of the RPC gap is tested for a duration of 48 hours. Multiple data samples are extracted from this large data sample and `$\textrm{C}_{\textrm{Leak}}$'s are calculated for each of them. The standard deviation of these estimated  `$\textrm{C}_{\textrm{Leak}}$'s, is calculated and treated as the systematic error. Figure~\ref{fig:systematic}(a) shows the distribution of the estimated  `$\textrm{C}_{\textrm{Leak}}$'s for data with duration of 25 hours. The relative systematic error for this case is 3.1\,\%. The relative systematic error for different lengths of duration of time is presented in Figure~\ref{fig:systematic}(b). As expected, it can be seen that with the increase in the length of the test duration, the relative systematic uncertainty in the measurement of $\textrm{C}_{\textrm{Leak}}$ decreases significantly.

\section{Result of Tested RPC Gaps}
\begin{figure}[h!]
  \vspace*{-15pt}
  \centering
  \includegraphics[width=0.99\textwidth]{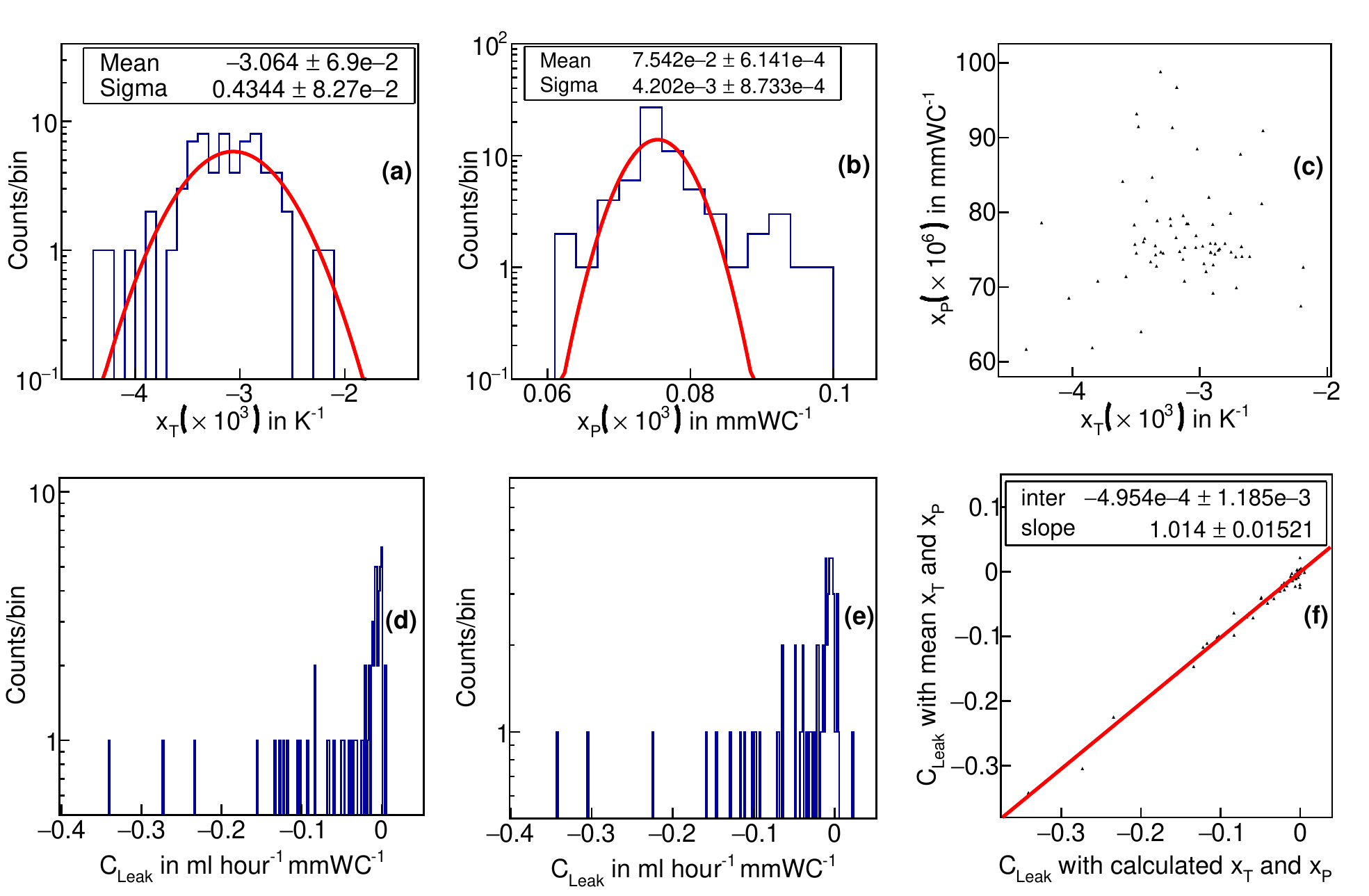}
  \caption{\textbf{(a)} $x_T$ values of eighty RPC gaps, \textbf{(b)} $x_P$ values of eighty RPC gaps, \textbf{(c)} Correlation between minimised values of $x_T$ and $x_P$ for eighty RPC gaps, \textbf{(d)} $\textrm{C}_{\textrm{Leak}}$ of eighty RPC gaps with calculated values of $x_T$ and $x_P$, \textbf{(e)} $\textrm{C}_{\textrm{Leak}}$ of eighty RPC gaps with mean values of $x_T$ and $x_P$, \textbf{(f)} Correlation between $\textrm{C}_{\textrm{Leak}}$ obtained by both calculated and mean values of $x_T$ and $x_P$ for eighty RPC gaps.}
  \label{fig:conclusion}
\end{figure}

The equipment and procedures described in this paper have been used to test eighty RPC gaps. The results of these tests are summarised in Figure~\ref{fig:conclusion} and general features are listed here.
\begin{itemize} \itemsep -5pt
\item The correction parameters depend on the structure of the RPC gaps. Values of $x_T$ and $x_P$ should be same for similar type of RPC gaps. Any gap with popped up button spacers(s) (`button pop-up' is discussed in Section \ref{sec:button}) will reflect higher values of these two parameters. This causes the spread in the Figure~\ref{fig:conclusion}(a) and \ref{fig:conclusion}(b).
\item Since, minimised values of $x_T$ and $x_P$ for a gas gap are two independent parameters, it can be seen in Figure~\ref{fig:conclusion}(c) that there is no correlation between these two parameters for eighty RPC gaps which were tested.
\item Figure~\ref{fig:conclusion}(d) shows the values of $\textrm{C}_{\textrm{Leak}}$ for eighty RPC gaps where the values of $x_T$ and $x_P$ are calculated explicitly for each RPC gaps. Figure~\ref{fig:conclusion}(e) shows the same but with the mean values of $x_T$ and $x_P$ obtained from \ref{fig:conclusion}(a) and \ref{fig:conclusion}(b).
\item Figure~\ref{fig:conclusion}(f) shows correlation between $\textrm{C}_{\textrm{Leak}}$ obtained by both calculated chamber by chamber and mean $x_T$ and $x_P$ values. The nature of correlation is direct. So, mean values of $x_T$ and $x_P$ can also be used for a quick calculation saving CPU time.
\end{itemize}

\section{Conclusion}
The method outlined above can give a quantitative estimation of the leak of an RPC in much less time compared to what may be obtained using conventional manometer. This method can be used for any kind of sealed chambers other than RPC where the structure of the chamber is prone to deform due to variation in ambient pressure and temperature. As discussed in reference \cite{Mondal_2016}, it is possible to eliminate the effect caused by ambient temperature by the use of a enclosure where the temperature is controlled precisely in case of abrupt and/or irregular changes in ambient temperature, causing difficulties in the calculation of volume correction factors. It should be noted that the minimal time required for a test will depend on the leakage. If the leak rate is very small then more time is needed and vice versa, but most of the gaps were able to test within 7-8 hours with a accuracy of $\sim 2\times 10^{-3}$\,ml\,hour$^{-1}$\,mmWC$^{-1}$. Currently, a gas gap is considered as usable if its $\textrm{C}_{\textrm{Leak}}$ value is greater than $-0.02$\,ml\,hour$^{-1}$\,mmWC$^{-1}$. If the CLS is operated at an excess pressure of 10\,mmWC above atmospheric pressure, the total leakage from ICAL detector will be approximately 6\,litres/hour during its active operation.

\end{document}